\documentclass[%
reprint,
nofootinbib,
bibnotes,
 amsmath,amssymb,
 aps,
pra,
]{revtex4-1}
\usepackage{natbib}
\usepackage{graphicx}
\usepackage{dcolumn}
\usepackage{bm}

\nocite{*}

\begin{document}

\preprint{APS/123-QED}

\title{ A new recurrences based technique for detecting robust extrema in long temperature records}

\author{Davide Faranda}
\affiliation{%
Laboratoire SPHYNX, Service de Physique de l'Etat Condens\`e, DSM,
CEA Saclay, CNRS URA 2464, 91191 Gif-sur-Yvette, France, Email: davide.faranda@cea.fr.
}
\author{Sandro Vaienti}%
\affiliation{%
Aix Marseille Universit\'e, CNRS, CPT, UMR 7332, 13288 Marseille, France and
Universit\'e de Toulon, CNRS, CPT, UMR 7332, 83957 La Garde, France, Email: vaienti@cpt.univ-mrs.fr.
}

\begin{abstract}  In this paper, by using new techniques originally developed for the analysis of extreme values   of dynamical systems, we analyze several long records of temperatures at different locations showing that they have the same recurrence time statistics of a chaotic dynamical system perturbed with dynamical noise and by instrument errors. The technique provides   a criterion  to discriminate whether the recurrence of a certain temperature belongs to the natural climate variability  or can be considered as a real extreme event with respect to a specific time scale fixed as parameter. The method gives a self-consistent estimation of the convergence.   \end{abstract}





\maketitle

\section{Introduction}

The definition of what is part of the natural variability of a system and of what is, instead, a proper extreme event is an evergreen topic among natural scientists for the effects these phenomena may have on social and economical  human activities  \cite{1,2,3,3a}. An extreme event is usually identified as an observation whose occurrence is unlikely with respect to a time scale of interest.  It is therefore natural to link the concept of extreme events to the recurrence statistics of certain observations.  Two families of techniques have been devised to tackle this problem:

{\em-Statistical based techniques} have been devised to estimate the tail  distribution by exploiting known properties of the bulk statistics   via the so called Extreme Value Theory (EVT) \cite{6,7}.  Extremes are extracted in a precise time window  and then fitted to the Generalized Extreme Value (GEV) distribution.  When the asymptotic distribution is known, one can  compute return times for any observations but this  is usually precluded for time series.  In this case the estimation  of return levels for very long return periods is prone to large sampling errors and  potentially large biases due to inexact knowledge of the shape of the tails  \cite{6}.  

{\em-Dynamical systems based techniques}  rely on defining extremes  via  a Poincar\`e recurrences analysis as points of the phase spaces visited only sporadically. They requires some knowledge of   the attractor underlying the observations which should been reconstructed  via the Ruelle-Takens embedding \cite{4}.   Unfortunately, to extract all this information usually requires the use of large data-sets since the convergence often depends by the dimensionality of the phase space and there is not an a priori way to recognize whether it is achieved or not \cite{5}.  

In this paper we present a formal way to define extreme events  based on a combination of the classical theory of Poincar\`e recurrences with the statistical results of the EVT.   By applying a special observable (see below) to the time series considered as the output of a stochastically perturbed dynamical system,  the asymptotic GEV parameters are known and the convergence does not depend on the dimensionality of the system but only on the time window fixed to extract the extreme events and on how fast a certain observation occurs.   A strong theoretical support for applying this technique has been given in \citep{8} by analyzing the convergence of the method  on series obtained as output of dynamical system   perturbed with instrument-like-error. We defer to \citep{8} for all the theoretical considerations and the proofs of the results whereas here we present the method in an easily reproducible algorithmic way. We will focus on some long records of  daily mean temperatures and the related temperature anomalies collected at different locations. Besides the intrinsic  interest of defining rigorous properties for temperature recurrences , this choice is also justified by the abundance and the good quality of the available data-sets.


\section{Recurrences as extreme events of dynamical systems}

The important concept of recurrences around a point of interest for the dynamics originate in the work of Poincar\'e \cite{9} and has been used trough the last century to study the properties of dynamical systems.
In the last year a growing attention has been reserved to the study of extreme events of observables originating from orbits of dynamical systems. In fact, even if the  Extreme Value Theory (EVT) has been devised for the study of independent and identically distributed (i.i.d.) variables, convergence towards the classical EVLs has been observed for chaotic dynamical system under special observables whose extremes are the recurrences around a point of the trajectory. We briefly recall some basics facts of the EVT, referring to the book by Leadbetter et al.  \cite{10} for further insights:
Gnedenko \cite{11} studied the convergence of maxima of i.i.d. variables $x(0),x(1), ..., x(s)$ sampled by applying the so-called block maxima procedure which consists in dividing the observations $x(t)$ into $n$ bins each containing $m$ observations and considering $M_j=\max\{ x(0),x(1), ..., x(m-1)\}, \ j=1,...,n$. He proved that, in the limit of $n,m \to \infty$ one gets as asymptotic distribution an EVL belonging to the Generalized Extreme Value (GEV) distribution family:
\begin{equation}
F_{G}(x; \mu, \sigma,
\xi)=\exp\left\{-\left[1+{\xi}\left(\frac{x-\mu}{\sigma}\right)\right]^{-1/{\xi}}\right\};
\label{cumul}
\end{equation}
which holds for $1+{\xi}(x-\mu)/\sigma>0 $ being  $\mu \in \mathbb{R}$ the location parameter, $\sigma>0$ the scale parameter and $\xi$ the tail index (also called shape parameter) discriminating the type of tails behavior: Gumbel EVLs  for bulk statistics with exponential tails $(\xi=0)$,  Fr\'echet  EVLs for fat unbounded tails $(\xi>0)$ and  Weibull EVLs for  upperly bounded tails $(\xi<0)$ .\\
In the last decade many works focused on the possibility of treating time series of observables of deterministic dynamical system  using  the EVT.
The first rigorous  mathematical approach goes back to the pioneer paper by Collet \cite{12}. Important contributions have successively been given in  \cite{13,14,15,16,17q,17,17b}.  The goal of all these investigations
was to find a suitable way to replace the independence conditions on the series $x(t)$ with the dependency structure introduced by the laws governing the output of dynamical systems. This is indeed possible by exploiting the properties of the Poincar\`e recurrences for chaotic  systems: it has been proved that if one considers a point of the phase space $\zeta$ and take as observable a function $y(t)=g(\mbox{dist}(\zeta, x(t)))$ i.e. the series of the distances between $\zeta$ and the other points of the orbits conveniently weighted by a function $g$, once sampled maxima of the observable $y(t)$, one gets asymptotic  convergence to one of the EVLs classical laws. In particular, if   $g(\cdot)=-\log(\cdot)$ is selected, the asymptotic EVL is always a Gumbel law with  shape parameter   $\xi=0$. Note that maxima of $g$ correspond to minima of the distance series thus, in each bin, we extract exactly the closest recurrence to the observation $\zeta$.

Here we adapt the method for finite time data-sets in a similar fashion of what has been done for the Lyapunov exponents by Wolf et al. \cite{18}. We   define the following algorithm for recognizing the points of the series around which there are few recurrences , provided that the series examined is chaotic:
\begin{enumerate}
\item Consider a given time series $ x(t)\ t=1,2,...,s$ .
\item Fix the point $\zeta$ to be a point of the series itself $x(t)$.
\item Compute the series $y(t)=-\log(\mbox{dist}( \zeta, x(t) ))$.
\item Once divided the series $y(t)$ in $n$ bins each containing $m$ data ($mn=s$), extract the maxima $M_j,\ j=1,...,n$ for the series $y(t)$.
\item Fit the maxima to the GEV model, perform a   Kolmogorov-Smirnov test \cite{a} to check whether the fit succeeded or failed.
\end{enumerate}
At this point    the results of the fit are compared to the output of a chaotic dynamical system perturbed by observational and dynamical noise.
\begin{itemize}
\item  If the fit succeeds we can repeat the experiment for shorter bin lengths and find the smallest $m$ such that, for the chosen $\zeta$, the fit converges. This defines the shortest convergent recurrence time.
\item  If the fit fails one should repeat the experiment by increasing the size of  $m$ until it is possible to retain a sufficient number of maxima to perform a reliable fit to the GEV model.
\end{itemize}

We can define the points such that the fit is non convergent as extreme of the series with respect to the bin length fixed by $m$  by exploiting the results given in \cite{17b,8}: if the point $\zeta$ is visited with less frequency, being the EVL parameters dependent on the  density of observations around the chosen $\zeta$ via the intensity of  the observational noise, one must go to higher values of $m$ in order  to have a reliable statistics. In \cite{8} the case of dynamical system perturbed with instrument-induced-error is treated in a formal way, by showing that classical EVls hold  for the orbit $\{T^ix\}_{i\ge 1}$ of a smooth and chaotic map $T$ in some manifold of dimension $d$ and preserving the invariant measure $\nu$  perturbed  with  observational noise.  Its effect is to change the orbit of the initial point $x$ into $T^ix+\varepsilon \psi_i$, where the $\psi_i$ are i.i.d. random variables which take values in the unit ball of  $\mathbb{R}^d$ and distributed according to the Lebesgue measure  $\theta$,  and $\varepsilon$ is a small positive parameter. The process $y(T^ix+\varepsilon \psi_i)$ is endowed with the probability stationary measure $\nu \times \theta^{\mathbb{N}}.$ Under some conditions on the choice of the map and of the measure, we have been able to prove that the previous process converge to a GEV distribution (Gumbel), with the scaling parameters $\mu$ and $\sigma$ given respectively  by
$\mu=\frac1d \ \log(\frac{K_d\ m \ \nu(B(\zeta,\epsilon)) }{\varepsilon^d})$ and $\sigma=\frac1d$, where $B(\zeta, r)$ denotes  the ball  centered at the point $\zeta$ and with radius $r$ and $K_d$ is the volume of the unit sphere in $\mathbb{R}^d$.  Whenever  the point $\zeta$ is visited with less frequency, the local density is  of lower order in  $\varepsilon$, which means that one should go to higher values of $m$ in order  to have a reliable statistics. Here we exploit   the  dependence  on $m$ to track the points visited with less frequency but one can use the  fact that instead classical EVL scaling constants do not depend on $m$ for studying other interesting properties of dynamical systems such as the determination of fractal dimensions \cite{8}.  We have already mentioned that one could also compute the recurrences after reconstructing the attractor via the traditional embedding techniques  \cite{4} by using the nearest neighbor techniques. In this case one has to estimate the maximum embedding dimension and the  time delay and after compute the recurrences. This is however infeasible due to the high dimensionality of the climate attractor as stated by Eckman and Ruelle \cite{19} and Lorenz \cite{20} . Our method does not need any computation of the embedding dimensions which appears only as a scaling factor  for the parameters $\sigma$ and $\mu$  but not on $\xi$ which must be zero in any case.  This motivates the choice of $g(\cdot)=-\log(\cdot)$  as the only observable whose GEV shape parameter does not depend on the dimension \cite{13}.
\section{Analysis of long temperature  records}

\begin{figure}
\centerline{\includegraphics[width=0.45449\textwidth]{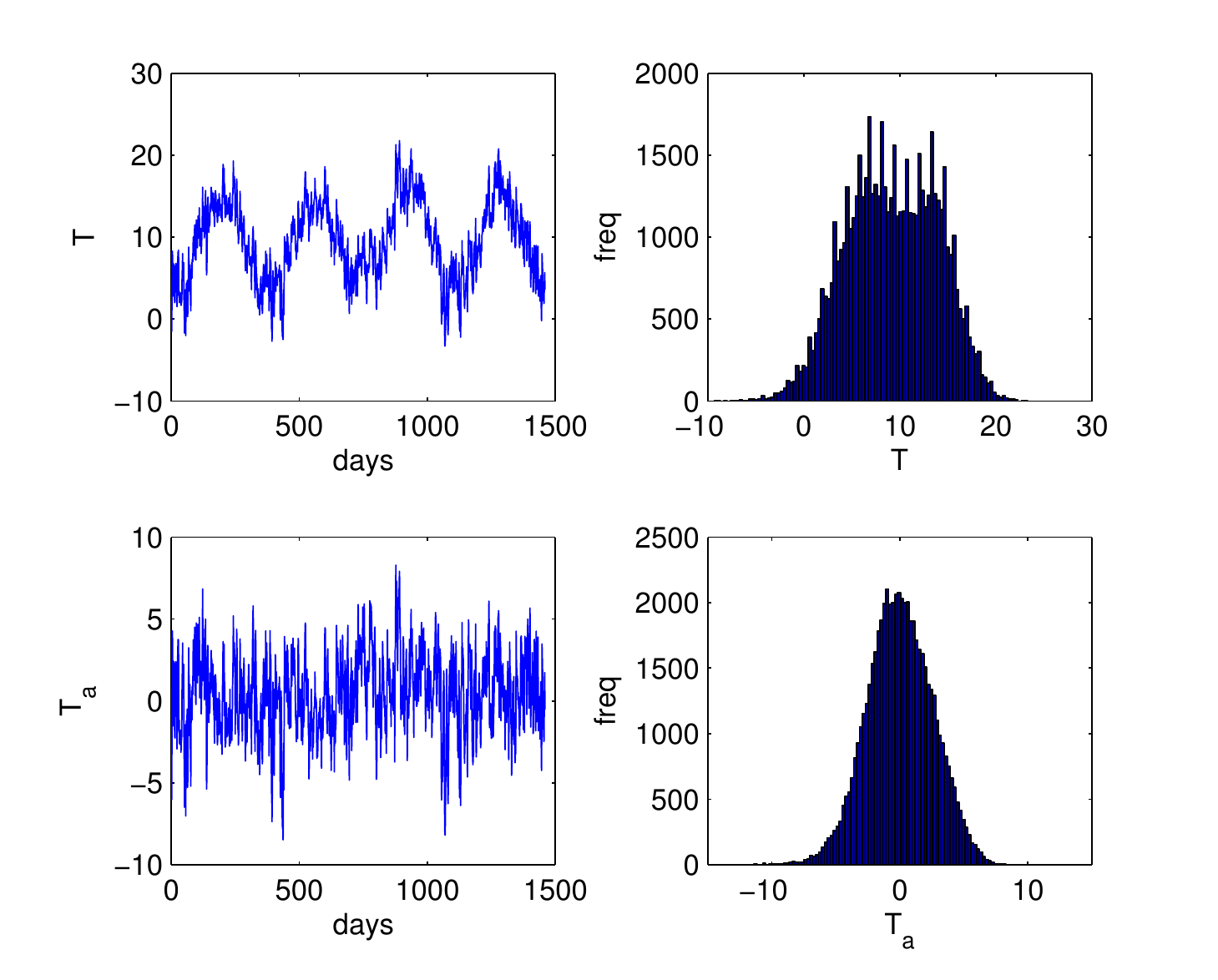}}
\caption{Top: An example of temperature series and its histogram. Bottom: the same for a temperature anomalies series. All the plots refer to   Armagh (UK) weather station.}
\label{serie}
\end{figure}

 At each location we will consider two different time series: the series of daily mean temperatures $T(t)$ and the series of daily  mean temperature anomalies $T_a(t)$. The latter has been obtained by removing from the data two annual cycles, computed with all data available percentiles in a 5-day window, in other words the series of anomalies is constructed by subtracting the best estimate of the seasonal cycle from the daily temperature. The series  have been all been extracted from the ECA\&D v1.1 database \cite{21}. An example of $T(t)$ and $T_a(t)$ with relative histograms is presented in Fig.~ \ref{serie}. The only assumptions on the time series analyzed concern their chaoticity and their stationarity. The first assumption is justified: \newline
\textit{-a priori}    by  observing that the series consist of a strong chaotic component - driven by the meteorological phenomena -  superimposed to a seasonal cycle. \newline
\textit{-a posteriori}  by  the successful application of our technique which would not work if the periodic component would dominate the chaotic one. \newline
For the stationarity of the series we instead refer to the World Meteorological Organization guidelines \cite{6}. We     report the analysis of three station chosen for their peculiar  climate characteristics: Armagh in Northen Ireland (UK), whose climate  is  influenced by the Atlantic Ocean with temperature ranges not extreme \cite{24}; Milan (Italy) whose climatology is governed by a  Mediterranean component and a more continental one due the proximity to the Alps and the Apennines mountains ranges \cite{25} and Wien (Austria) whose climate has already marked continental features \cite{26}. The series feature respectively 161, 246, 156 years of  observations. In order to perform a reliable GEV fit, we have to take into account the role of the truncation error being the series of temperature  truncated at $p=3$ digits.  As discussed in \cite{8},  a blind application of the method without recovering a continuous temperature field, would lead to a divergent fit as the recurrence distribution will appear as a collection of Dirac's delta. We can solve the problem  by adding a  uniform noise  for $p\geq4$ which does not alter the observations but allows for recovering continuity.\\



\begin{figure}
\centerline{\includegraphics[width=0.46449\textwidth]{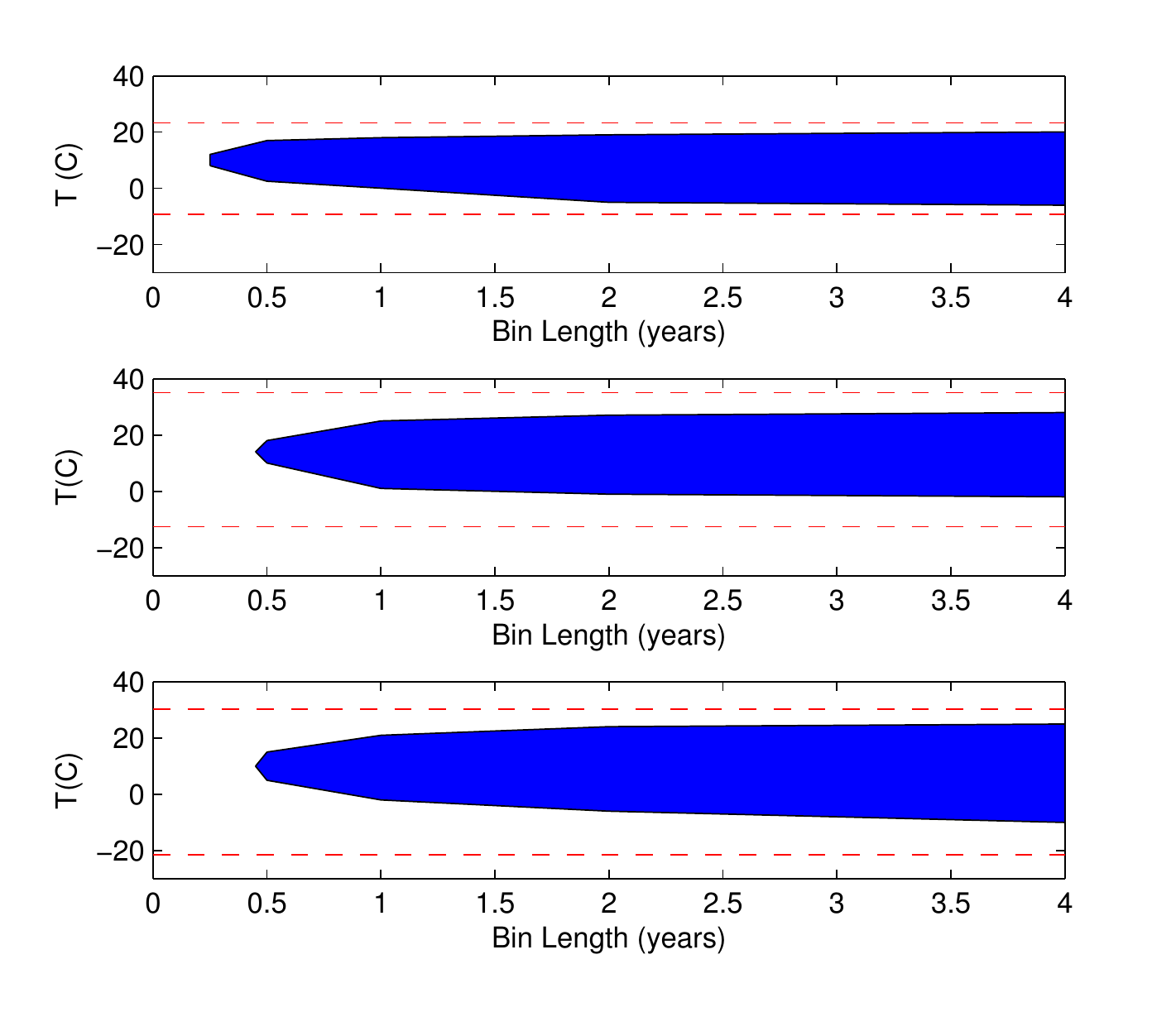}}
\caption{ Region of temperature with convergent Gumbel fit (blue area) for different bin length. Red dotted lines: absolute extremes of the temperature series for  Armagh (top), Milan (middle) and Wien (bottom).  Extremes are located in the white area of the plots between the two dotted lines.}
\label{Trange}
\end{figure}
\begin{figure}
\centerline{\includegraphics[width=0.4999\textwidth]{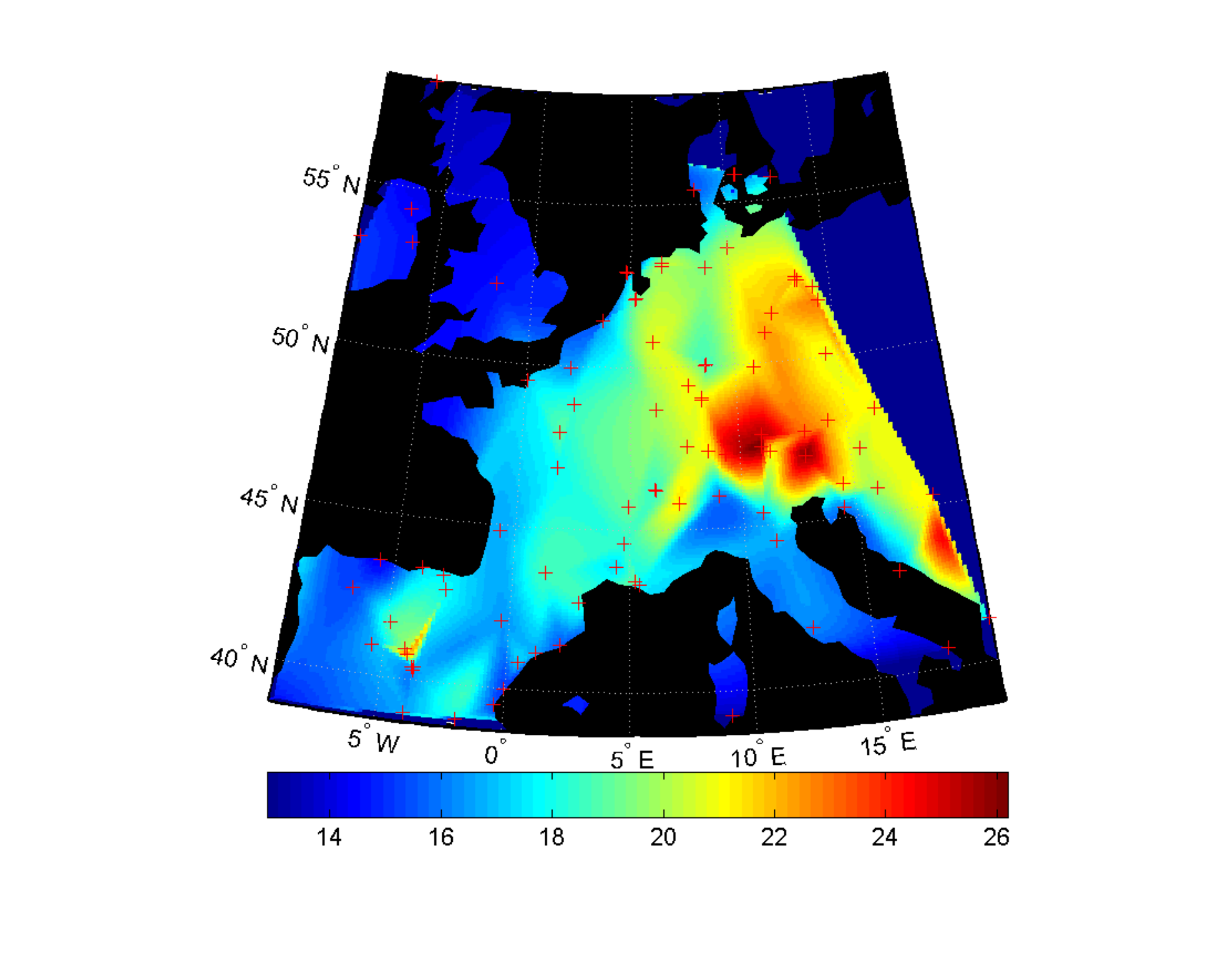}}
\caption{ Map of the range of \textit{admissible anomalies} for the European region, obtained considering the interval of temperature anomalies $\zeta$ such that the fit passes the Kolmogorov Smirnov test. The red crosses represent the location of the stations used for the analysis.}
\label{tang}
\end{figure}

The shape parameter $\xi$  has been obtained  by fitting the GEV distribution via a maximum likelihood estimation technique \cite{17q}. The recurrences, computed for all the temperatures     $\zeta$   between the absolute recorded extremes, is presented in Fig~\ref{Trange} for  the stations located Armagh (top panel), Milan (middle panel) and Wien (bottom panel). The experiments have been repeated for different bin lengths between 3 months and 4 years. In  Fig~\ref{Trange}, the blue area represents the reference temperatures $\zeta$ whose GEV   fit passes the Kolmogorov Smirnov test to the Gumbel law \cite{a}. This temperatures range is what we propose as a rigorous definition of natural climate variability with respect to the time scale given by $m$.  Real extreme events are instead located outside the white blue area and correspond to unsuccessful Gumbel fits.  For bin length shorter than six months  the fits fail at any reference temperatures as the bin length  $m$ is too short for observing proper recurrences  near $\zeta$.  The only exception is registered  at Armagh where, due to the  limited seasonal temperature excursions,  the convergence is achieved for $8 C <\zeta < 12 C$ already for a 3 months bin length.  In general one observe that, when the bin length is increased, the range of temperatures accepted as natural climate variability increases e.g. for Milan one could define $0$ C as an extreme temperature  with respect to a   bin length of half a year but not when considering a bin length of 4 years. We repeated  the same analysis reported in Fig \ref{Trange} for the series of the anomalies $T_a$. The main advantage of using temperature anomalies consists in the possibility of comparing the  climatology of different locations. Let us consider  as bin length a one year period: at Armagh, in a temperate, marine climate, only anomalies up to $\pm$ 6 C are acceptable in the annual variability according to the method described. At Wien, the continental climate fosters large temperature excursions so that we find  the anomalies up to $\pm$ 10 C  acceptable in the annual variability. For Milan we get $\pm 7$ C,  an intermediate situation which recalls the characteristics of a climate influenced both by the Mediterranean sea and by its location in the Po valley.  By plotting the length of the interval of convergent $T_a$ at all the European locations for which   at least 60 years of daily data  are available, we can construct a climatological map of Europe.  The results are reported in Fig \ref{tang} for the bin length of one year.  Different climatic regions are well highlighted: the British Isles, Brittany , Italy and the coastal areas of the Iberic peninsula   have milder climate with a  significantly lower range of admissible anomalies. The highest excursions are admissible in continental Europe and in correspondence of mountains range. This explains the gradient in the central Iberic peninsula: one of the station, Navacerrada, is 1800  high and its admissible excursions range over 24 degree.

\section{Final Remarks}

The main achievement of this paper is to suggest a method for discriminating between real temperature extremes and natural climate variability  by analyzing the recurrence properties of a given temperature at a certain location.   The recurrence technique shows that the concept of natural climate variability is well associated to a precise time scale here directly controllable by tuning $m$.   We can define the minimum return time around a certain temperature chosen as $\zeta$, the smallest $m$ such that the method converges. In our analysis  we  can retain a sufficient number of maxima only for bin length shorter than  4 year  so that   we cannot study extremes of longer time intervals but we hope that the method will  be tested to  longer time series   e.g. climate models output. The applicability of this technique, originally developed for studying chaotic and low dimensional dynamical systems, relies on a careful analysis on the role of the noise introduced by the instrument which cannot be ignored in the study of time series.     Moreover, by analyzing the anomalies of about 100 stations we can easily construct climatological maps of anomalies where differences between continental and mild, maritime climates are well highlighted. The method itself gives an estimate of the convergence in terms of deviation by a Gumbel law and, once the stationarity and the chaoticity of the series can be assumed,   does not require any the computation of other properties of the series such as the embedding dimension .  The method  applicability is currently being tested on turbulence data-sets produced in laboratory experiments with encouraging results that will be reported in a future publication.








\medskip\noindent\textbf{Acknowledgement}.
SV was supported by the ANR-
Project {\em Perturbations}, by
the PICS ( Projet International de Coop\'eration Scientifique), Propri\'et\'es statistiques des sys\`emes dynamiques det\'erministes et al\'eatoires, with the University of Houston, n. PICS05968 and by the projet MODE TER COM supported by {\em R\'egion PACA, France}.  SV thanks support from  FCT (Portugal) project PTDC/MAT/120346/2010. The authors aknowledge Paul Manneville, Berengere Dubrulle and Emmanuel Virot for useful discussion and suggestions on the paper.

\end{document}